\documentclass[journal,comsoc]{IEEEtran}
\usepackage[utf8]{inputenc} 
\usepackage[T1]{fontenc}
\usepackage{url}
\usepackage{ifthen,hyperref}
\usepackage[cmex10]{amsmath}
\usepackage{array} 

\usepackage{enumerate}
\usepackage{amssymb}
\usepackage{amsmath} 

\usepackage{algorithm}
\usepackage{algorithmicx}
\usepackage{algpseudocode} 

\usepackage{hhline}
\usepackage{mathrsfs}
\usepackage{bm}
\usepackage{tikz}
\usetikzlibrary{arrows}
\usepackage[caption=false]{subfig}
\usepackage{graphicx,booktabs,multirow}
\usepackage{epstopdf}
\usepackage{multirow}
\usepackage{tabularx} 
\usepackage{booktabs}
\usepackage{cite}

\usepackage{pifont}

\definecolor{colorhkust}{RGB}{20,43,140}
\definecolor{colortsinghua}{RGB}{116,52,129}
\definecolor{color1}{RGB}{128,0,0}

\usepackage{amsthm}

\theoremstyle{definition}

\theoremstyle{remark}

\begin{document}
      \title{WirelessAgent: Large Language Model Agents for Intelligent Wireless Networks} 
      \author{Jingwen Tong, Jiawei Shao, Qiong Wu, Wei Guo, Zijian Li, Zehong Lin, and Jun Zhang,~\IEEEmembership{Fellow,~IEEE}
      \thanks{This work was supported by the Hong Kong Research Grants Council under the Areas of Excellence scheme grant AoE/E-601/22-R and NSFC/RGC Collaborative Research Scheme grant CRS\_HKUST603/22. 
     The authors are with the Department of Electronic and Computer Engineering, Hong Kong University of Science and Technology, Hong Kong.  (The corresponding author is J. Zhang). }}
        \maketitle
\begin{abstract} 
Wireless networks are increasingly facing challenges due to their expanding scale and complexity. 
These challenges underscore the need for advanced AI-driven strategies, particularly in the upcoming 6G networks. 
In this article, we introduce \textit{WirelessAgent}, a novel approach leveraging large language models (LLMs) to develop AI agents capable of managing complex tasks in wireless networks.
It can effectively improve network performance through advanced reasoning, multimodal data processing, and autonomous decision making.
Thereafter, we demonstrate the practical applicability and benefits of WirelessAgent for network slicing management.
The experimental results show that WirelessAgent is capable of accurately understanding user intent, effectively allocating slice resources, and consistently maintaining optimal performance.

\end{abstract}


\section{Introduction}
Wireless communications have become the cornerstone of modern society, profoundly impacting daily lives and driving innovations across industries. As human society develops and technology advances, wireless networks face unprecedented complexity due to their expanding scale, density, and technological diversity. Traditional optimization and machine learning approaches are increasingly inadequate for managing these challenges, prompting a shift towards more advanced artificial intelligence (AI) in 6G networks \cite{letaief2019roadmap}. 

Recently, it has become a consensus to enhance the intelligence of 6G networks to manage the increasing network complexity, support AI-driven applications, and achieve unparalleled performance \cite{bariah2024large}. 
However, existing AI solutions for wireless communications are often problem-specific and lack generalizability. 
This limitation underscores the urgent need to develop more versatile and generalizable AI algorithms to address a wide range of wireless challenges, moving towards true intelligence in the field \cite{shao2024wirelessllm}. 
This need has given birth to the concept of AI agents in wireless communications, leveraging recent advancements in large language models (LLMs) to automate various wireless tasks \cite{brown2020language}.

With the release of ChatGPT, LLMs have attracted significant attention and greatly impacted various research areas due to their human-like text comprehension and generation.
Besides, they have showcased their potential in a range of cognitive capabilities when applied to AI agents, making a significant expansion beyond natural language processing into reasoning and decision making.
In addition, the capabilities of LLMs have been further pushed by the advent of multimodal generative AI systems, such as GPT-4o, which enhances the capabilities of LLMs by incorporating visual, auditory, and other non-textual information.
Moreover, the integration of external knowledge bases grounds LLMs, allowing them to leverage domain-specific information for more accurate and context-aware responses.

While LLMs have demonstrated remarkable capabilities in text understanding and generation, there is a natural gap between general-purpose language tasks and practical applications in wireless networks. 
Many studies on LLMs are limited to a direct application for telecom language understanding \cite{erak2024leveraging, maatouk2024large, dandoush2024large}. 
The full capabilities of LLMs as intelligent and autonomous agents to manage and optimize network operations remain underexplored, particularly in their ability to interpret wireless data, decompose complex tasks, adapt to changing conditions, and utilize external resources.
This article aims to unlock the potential of LLMs for AI agents in wireless networks. Our main contributions are summarized as follows:
\begin{itemize}
    \item We put forth \textit{WirelessAgent}, a framework that empowers LLMs with four core modules: perception, memory, planning, and action. 
    WirelessAgent aims to interpret multimodal input, automate complex tasks accordingly, and output solutions with the assistance of external knowledge bases and powerful tools.

    \item We provide a proof-of-concept case study for network slicing management, which demonstrates the effectiveness of WirelessAgent in accurately understanding user intent, effectively allocating slice resources, and consistently maintaining optimal performance.

    \item We outline potential future research directions for WirelessAgent, including multimodal integration, privacy and security concerns, and real-world deployment and evaluation.
\end{itemize}

\section{Agents for Wireless Networks}
The integration of LLMs into the domain of wireless communications for developing wireless-domain agents represents a significant advancement in creating intelligent services and enhancing the performance of wireless networks.

\begin{figure*}[!t]
    \centering    \includegraphics[width=0.98\textwidth]{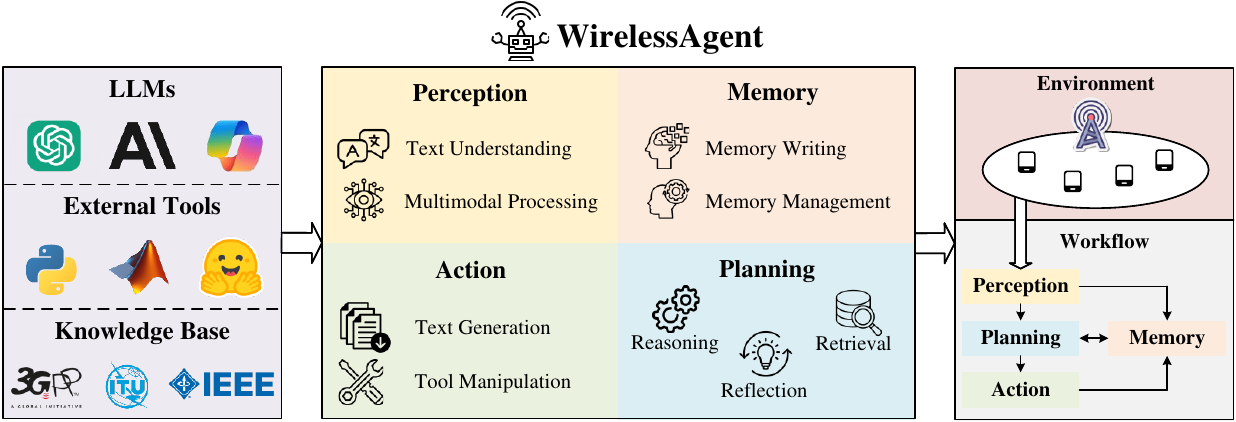}
    \caption{The overview of the WirelessAgent framework. From left to right, the three parts are key supports, core modules, and basic usage.} \label{Fig:WirelessAgent}
\end{figure*}

\subsection{From LLMs to AI Agents}
LLMs have become essential in understanding and generating human-like text from provided input. This attribute is fundamental for developing AI agents capable of interpreting and responding to complex instructions from humans and other agents \cite{xi2023rise}.
In addition, LLMs serve as the cognitive core for AI agents, enhancing their functionality through techniques such as multimodal perception and tool utilization. 
This broadens their operational scope across various domains. 
Moreover, the capacity of LLMs for few-shot and zero-shot generalization allows for flexible adaptation to new tasks without extensive retraining.
The promising capabilities of AI agents underscore their potential utility in wireless communications, facilitating more intelligent and autonomous network systems \cite{xu2024large}. 
In the following, we explore the principles that guide the development of AI agents for wireless networks.

\subsection{Principles of Agents for Wireless Networks}

\textbf{Interaction.} An agent should be capable of effectively communicating and interfacing with humans, the environment, and other agents.
This interaction should be adaptive, allowing the agent to understand and respond to different communication modulations. 
Besides, agents in wireless networks should be able to interface with various wireless systems and protocols, enabling them to gather information, control parameters, and optimize performance based on the specific requirements of each wireless scenario.
Moreover, these agents should support collaboration among agents to improve decision making and resource management by sharing sensory data, computational resources, and network bandwidth, enabling more efficient task execution in heterogeneous networks.

\textbf{Autonomy.} 
An effective agent must function autonomously without direct intervention from humans.
It should have significant control over its actions and internal states.
Besides following explicit human instructions for task completion, it should be able to initiate proactive strategies and complete tasks without detailed step-by-step instructions.
In addition, an agent for wireless networks should be able to respond rapidly to immediate changes and even failures in telecommunication systems.
Without timely interventions, it needs to figure out effective countermeasures and adapt its operational strategies to avoid disruptions.

\textbf{Self-improvement.} The continual learning and adaptation capability is crucial for maintaining the relevance and effectiveness of an agent in the evolving field of wireless communications. 
An agent should incorporate mechanisms to learn from interactions, feedback, and environmental changes.
This involves updating its knowledge base based on new data and improving its capabilities to handle emerging challenges in the wireless domain.
Advanced AI techniques, such as prompt engineering and reinforcement learning, can enable agents to evolve over time, enhancing their intelligence and utility.

\section{The WirelessAgent Framework}
Following the principles of agents for wireless networks, we introduce \emph{WirelessAgent}, a framework leveraging LLMs to develop AI agents capable of managing complex tasks. Fig. \ref{Fig:WirelessAgent} is an overview of the WirelessAgent framework, which consists of the key supports, core modules, and basic usage. The WirelessAgent employs a structured workflow that emulates human sensory and response systems. The perception module, akin to sensory organs, comprehends natural language and translates multimodal wireless data for the agent. 
The memory module, acting as the hippocampus, stores and retrieves past interactions and learned behaviors.
The planning module, which mirrors the prefrontal cortex, processes information and makes decisions.
Finally, the action module, analogous to limbs, executes tasks using various tools, altering the environment. 
This cyclical process enables continuous feedback and dynamic interaction with the external world.
In the following, we elaborate on these three modules.

\subsection{Perception}
The perception module is adept at processing and understanding diverse forms of input, mirroring the capabilities of human sensory organs. It includes two key functionalities:

\subsubsection{Text understanding}
Language serves as a rich medium for communication, encapsulating extensive information.
Leveraging the advanced capabilities of LLMs, agents can proficiently engage in multi-language understanding and exhibit in-depth comprehension abilities.
During interactions between users and WirelessAgent, textual instructions are provided to LLM agents, including explicit requests and implied intentions.
In addition, by fine-tuning language models with specific datasets and professional corpus, WirelessAgent can interpret complex terminologies used in wireless communications, bridging the gap between technical language and user-friendly explanations.

\subsubsection{Multimodal processing}
WirelessAgent can autonomously perceive the surrounding environment using equipped sensors and collect multimodal data, encompassing 2D/3D vision and radio signals \cite{yin2023survey}.
Although LLMs exhibit outstanding performance in language conversations, they cannot inherently analyze multimodal data. 
The multimodality contains a wealth of information, including properties of objects, spatial relationships, and wireless channel conditions.
Such rich information offers the agent a broader context and a more precise understanding, deepening the perception of the environment.
To help LLMs process multimodal data, a straightforward approach is to generate corresponding text descriptions.
This approach is highly interpretable and does not require an additional training phase.

\subsection{Memory}
The memory mechanism empowers WirelessAgent to comprehensively analyze past and current data, enhancing its ability to manage dynamic information in wireless intelligence applications.
After the observations are perceived, a part of them will be stored by the agent for further usage through the memory writing operation.
Besides, the past mistakes, successful interventions, and learned behaviors derived from these experiences are recorded for future reference.
For wireless applications, when a user reports a connectivity issue in a specific location, WirelessAgent stores details like the user's identity (ID) and location, channel state information (CSI), network conditions, and the troubleshooting steps. 
This information is then structured and indexed for future access.
In addition, organizing and indexing historical records help in efficient data retrieval and reduce the memory footprint.

\subsection{Planning}
Planning is a key management function that facilitates complex task completion by organizing thoughts, outlining steps, and monitoring progress.
Typically, it involves three key modules: a reasoning module, a retrieval module, and a reflection module.

\subsubsection{Reasoning module}
Reasoning is crucial in human intellectual activities such as problem solving, decision making, and critical analysis. 
Similarly, for LLM agents, reasoning is essential for addressing complex tasks. 
They should break down complex tasks into manageable sub-tasks and formulate corresponding strategies.
The representative techniques that empower WirelessAgent to perform reasoning include in-context learning (ICL) and chain-of-thought (CoT) prompting.
ICL leverages the demonstrations provided within a prompt and analyzes the information presented in the immediate context to generate responses. 
Given the current state and parameters of a network, such as signal strength, noise levels, and user mobility, ICL can predict potential points of failure or recommend adjustments to optimize throughput and stability.
Besides, CoT reasoning explicitly prompts LLMs to generate intermediate steps or reasoning paths.
In dynamic spectrum management, by sequentially considering various constraints such as quality of service (QoS) requirements, the available resources, and interference levels, WirelessAgent can outline a logical pathway that leads to optimal spectrum utilization strategies.

\subsubsection{Retrieval module}
Retrieving the most appropriate content from either its internal memory or an external knowledge base is crucial for WirelessAgent to enhance the response quality.
Specifically, retrieval-augmented generation (RAG) \cite{RAG_lewis2020retrieval} gives LLMs access to information beyond their training data, retrieving extra sources to ground language models on the most relevant and up-to-date information.
The antecedent experience stored in the internal memory serves as a rich source of learned behaviors and previously encountered scenarios.
This internal knowledge allows the model to quickly adapt to similar situations in future interactions.
Besides, external knowledge integration is essential in fast-evolving fields and dynamic environments.
When queried about the latest 5G standards, the agent will prioritize retrieving the most recent information from 3GPP technical specifications and white papers over older documents.
It will also consider the relevance of each document to the specific aspect of 5G being inquired about, such as the physical layer, network architecture, or security features.

\subsubsection{Reflection module}
A reflection module is incorporated into WirelessAgent to emulate the cognitive learning process inherent in human decision making.
The reflection is divided into reflection-before-action and reflection-after-action manners, each serving distinct purposes to enhance the decisions. 
The reflection-before-action manner is thinking through a situation before making decisions or taking actions.
It involves reflecting on the relationship between the observations and the resultant wireless network performance, drawing connections between the provided information and the outcomes. 
The reflection-after-action manner analyzes the outcomes after an incident or an action. 
It examines past decisions, tracking actions and the subsequent results to learn from past successes or mistakes.
By purposefully designing prompts, WirelessAgent reflects on past network management decisions.

\subsection{Action}
In constructing WirelessAgent, the action module carries out specific commands to interact with the environment.

\subsubsection{Text generation}
The advancement of Transformer-based generative models has equipped LLM agents with language generation capabilities.
However, these language models tend to hallucinate or generate text that is fluent and plausible but factually incorrect.
This hallucination problem can be especially concerning when applying LLMs to domains like wireless communications, where consistency with the underlying physical principles and system constraints are critical.
There are many approaches to mitigate the hallucination issue and make WirelessAgent better follow instructions.
For instance, fine-tuning LLMs on wireless datasets such as technical publications, patents, and standards can significantly reduce the frequency of hallucinations. 
Besides, alignment techniques like reinforcement learning from human feedback can be employed to further train the models based on specific user interactions and corrections.

\subsubsection{Tool manipulation}
Tools enhance the capabilities of their users. 
When confronted with complex tasks, humans use tools to simplify the process and boost efficiency, which saves time and resources. While LLMs have extensive knowledge from the training data, they can sometimes misinterpret ambiguous prompts or even generate hallucinations.
Specialized tools, such as the Python, Matlab, and Huggingface platforms, help LLMs improve their performance, adapt to specific domains, and meet the unique requirements of those domains in a modular way \cite{shen2024hugginggpt}. 
For example, propagation prediction software, which uses the computationally efficient ray-tracing algorithm in Matlab software, generates accurate and explainable simulations of electromagnetic wave behavior in various environments.
This is crucial in fields like telecommunications, where understanding signal propagation can directly impact the design and optimization of networks.

\begin{figure*}[!t]
    \centering
    \includegraphics[width=0.98\textwidth]{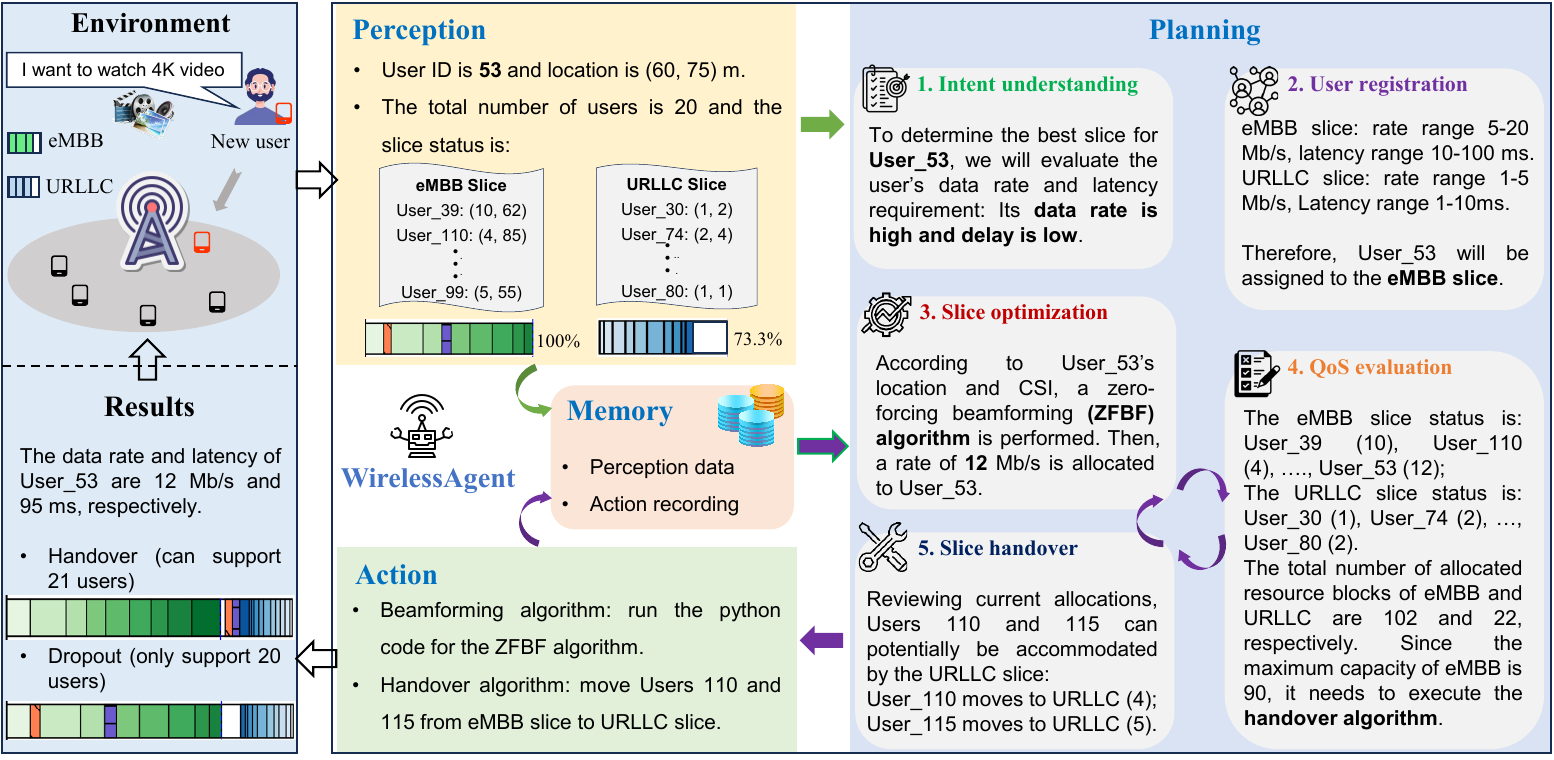}
    \caption{An illustration of WirelessAgent-enabled slice management, with eMBB and URLLC slices. This framework consists of two parts. The first part is the external environment, including human interaction and network conditions. The second part is the WirelessAgent, where different modules are related to different functions in the network slicing management task. In addition, an example of User 53 is used to visualize the agent's workflow.} \label{Case_Study}
\end{figure*}
\section{Case Study: network slicing Management}

This section provides a proof-of-concept case study for network slicing management built on WirelessAgent. We first introduce the fundamental concepts of network slicing and a traditional slicing architecture. We then put forth an intelligent slice management approach utilizing WirelessAgent. Finally, we conduct simulations to evaluate the proposed approach. 

\subsection{Network Slicing}
Network slicing is a key concept in 5G and beyond, designed to provide tailored network services to meet diverse requirements \cite{zhang2019overview}. 
It enables multiple virtual networks (or slices) to coexist on a shared physical infrastructure. There are three main types of network slices:
\begin{itemize}
\item \emph{Enhanced Mobile Broadband (eMBB)}: These slices focus on high data rates and capacity.
\item \emph{Ultra-Reliable Low-Latency Communications (URLLC)}: This type provides extremely low latency and high reliability, ideal for applications like autonomous vehicles.
\item \emph{Massive Machine-Type Communications (mMTC)}: These slices support many low-power, low-data-rate IoT devices.
\end{itemize}
Each slice type is tailored to specific requirements, ensuring optimal application performance.

The network slicing architecture usually consists of two main components: slice implementation and slice management. The first component is a multi-tier architecture with three layers: the service layer, the network function layer, and the infrastructure layer. Each layer has specific tasks that contribute to slice deployment and implementation. The second component is a centralized entity known as the network slice controller. This controller monitors and manages interactions among the three layers, ensuring efficient coordination and coexistence of multiple slices. 

Many studies are dedicated to developing network slicing management strategies in the second component \cite{li2018deep, zhang2017network}. Unfortunately, these methods often struggle with the complexity and dynamic nature of wireless networks. This phenomenon becomes more serious in dense networks. In this article, we put forth a WirelessAgent-enabled network slicing management approach, aiming to replace the traditional network slice controller with an intelligent agent.
 
\subsection{WirelessAgent-enabled Network Slicing Management}
The structure of the WirelessAgent-enabled slice management approach is shown in Fig. \ref{Case_Study}, which consists of two parts. The first part is the external environment, including human interaction and network conditions. The second part is the WirelessAgent, where different modules are tailored for the network slicing management task. Following the workflow of WirelessAgent, we explain how these modules collaborate to address the network slicing management task in the following.

\textbf{Perception:} The perception module effectively processes and understands two types of input information with the capabilities of text understanding and multimodal processing. The first type includes user features, such as the user's ID, location, CSI, and specific requirements. The second type involves network conditions, such as the number of users and the status of different slices. For example, as shown in Fig. \ref{Case_Study}, a new user comes to the cellular network and sends a request to the BS. The WirelessAgent perceives that the new user's ID is 53 and its location is $(60, 75)$ m. In addition, it gathers information on the total number of users in the cellular network and the status of eMBB and URLLC slices (e.g., the current resource occupation rates of eMBB and URLLC slices are $100\%$ and $73.3\%$, respectively). At last, this information is stored in the memory module and forwarded to the planning module for further processing.

\textbf{Memory:} The memory module stores and manages two main types of information: perception data and action recordings. First, perception data identify various states of the external environment, enabling the WirelessAgent to respond quickly when a new user is recognized in a known state. Second, action recordings help the WirelessAgent reflect on its behavior and avoid repeating mistakes.

\textbf{Planning:} The planning module is crucial in the WirelessAgent workflow, facilitating network slicing management by organizing tasks, outlining steps, and monitoring progress using the capabilities of reasoning, retrieval, and reflection. As shown in Fig. \ref{Case_Study}, WirelessAgent divides the network slicing task into five sub-tasks: intent understanding, user registration, slice optimization, QoS evaluation, and slice handover. Next, we elaborate on these steps using the example of User 53.

\begin{itemize}
    \item \emph{Intent understanding}: This step identifies the user's service requirements. For example, if User 53 wants to watch a 4K video, resources with a high data rate and low latency should be allocated. This step recommends a data rate range of $[12, 15]$ Mb/s and a latency of $90$ ms for User 53, utilizing the in-context learning technique. 
    \item \emph{User registration}: Based on the recommendations and slice decision boundaries, this step assigns the user to an appropriate slice. For instance, User 53 is assigned to the eMBB slice, as it meets the conditions of a data rate $[5, 20]$ Mb/s and latency $[1, 5]$ ms. 
    \item \emph{Slice optimization}: This step determines the user's data rate and latency according to its CSI and the slice's status using the beamforming algorithm. For User 53, the zero-forcing beamforming (ZFBF) algorithm is applied, assigning a data rate of 12 Mb/s. This involves retrieving internal memory and using external tools.
    \item \emph{QoS evaluation}: This step checks if all user requirements are met with the allocation results. If not, WirelessAgent adjusts resources and may reassign users to different slices during the slice handover step; otherwise, the optimization results are finalized. For example, if the eMBB slice cannot support 21 users as User 53 requires a higher data rate, adjustments are needed.     
    \item \emph{Slice handover}: Based on evaluation results, this step reassesses all users' requirements and reallocates them as needed. For instance, Users 110 and 115 might be moved from the eMBB slice to the URLLC slice to accommodate User 53's data rate needs. After the handover, resources must be reallocated, prompting a return to the slice optimization and QoS evaluation steps. This cyclical process demonstrates the agent's reflection ability.      
\end{itemize}

\textbf{Action:} The action module executes commands from the planning module to interact with both internal and external environments. It generates text to communicate with humans, ensuring that their requirements are understood, and utilizes tools to enhance capabilities. For instance, during the slice optimization step, the action module selects different beamforming and handover algorithms to maximize network performance. Finally, it outputs the allocation results for all users.   

In summary, the proposed structure employs two workflows to handle the network slicing management task corresponding to the principles of WirelessAgent, as shown in Fig. \ref{Case_Study}. The main workflow is from the external environment to perception, planning, action, and the external environment. This workflow ensures that the WirelessAgent can fulfill the network slicing task following the principles of \emph{interaction} and \emph{autonomy}. Another workflow is from planning to action, memory, and planning. This workflow enables WirelessAgent to \emph{self-improve}, enhancing its intelligence and utility. In the following, we show that WirelessAgent can efficiently handle the network slicing management tasks by following these two workflows.

\subsection{Numerical Results}

\begin{figure}[!t]
    \centering    \includegraphics[width=0.48\textwidth]{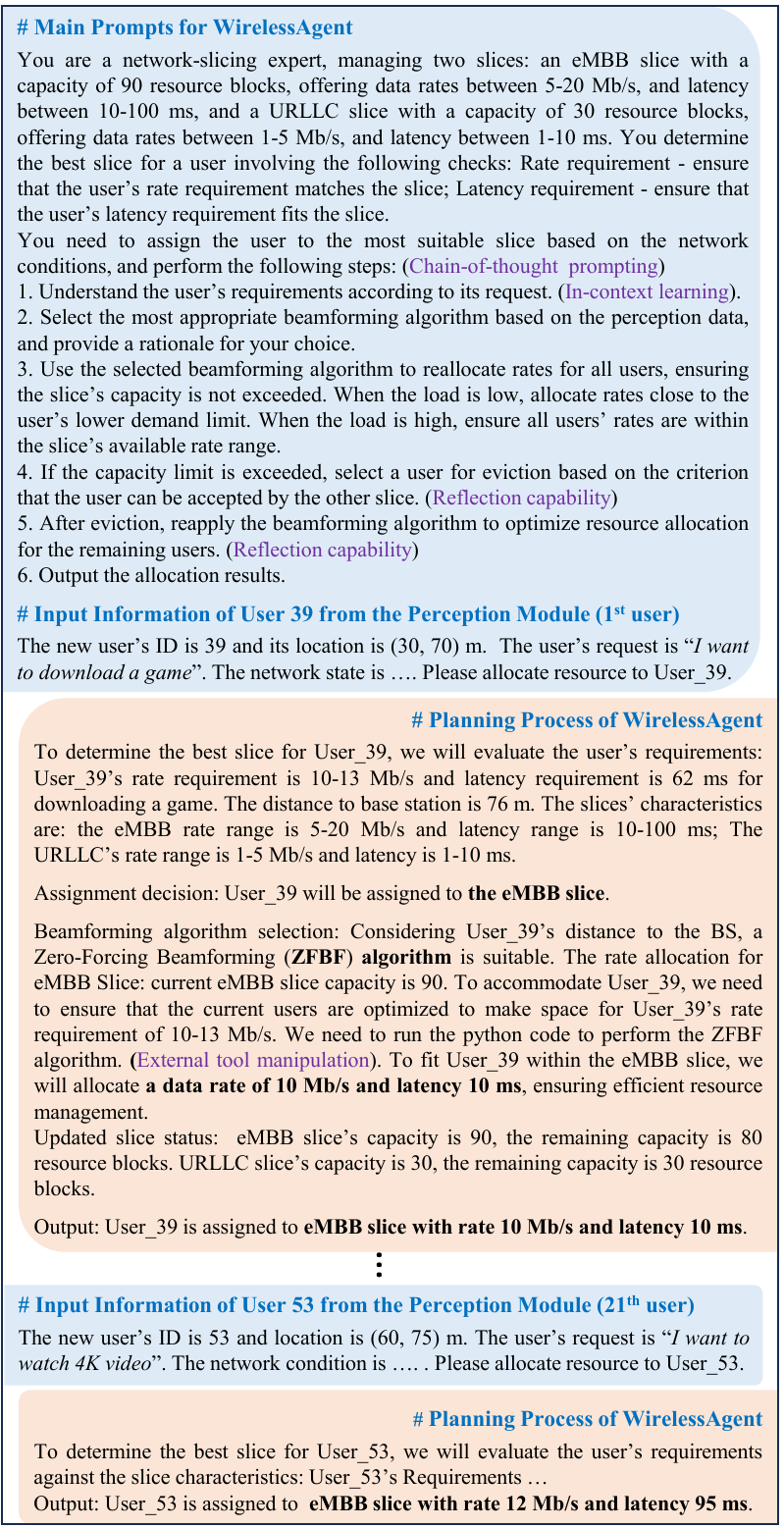}
    \caption{An example of WirelessAgent used in the network slicing management task.} \label{Fig:ImpleDetails}
\end{figure}

\begin{figure*}[!t]
\centering
\subfloat[The traditional network slicing management approach]{\label{TradM}
\includegraphics[width=0.95 \columnwidth]{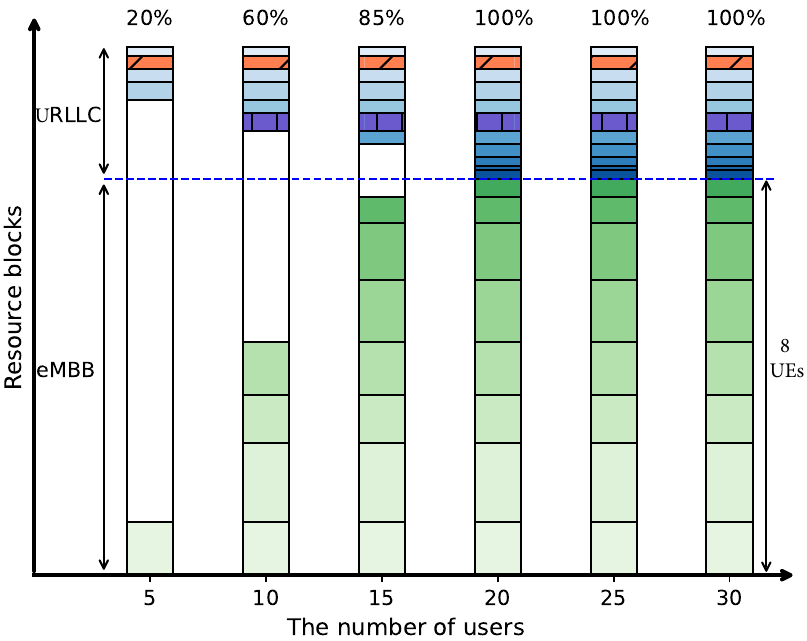}}
\hfil
\subfloat[The WirelessAgent-enabled network slicing management approach] {\label{WirelessAgentM}
\includegraphics[width=0.95 \columnwidth]{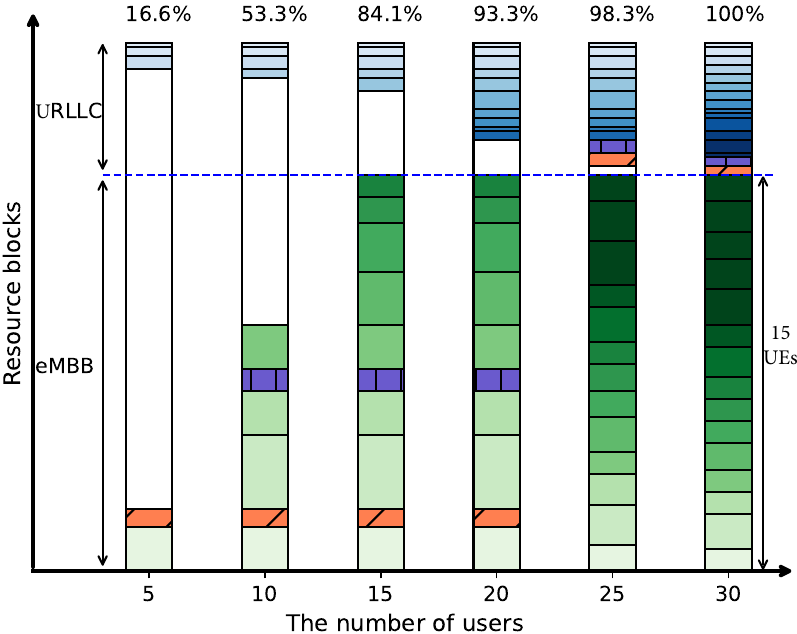}}
\caption{The resource occupation rate of the traditional and WirelessAgent network slicing management approaches in a cellular network, where users arrive sequentially. The total number of RBs in the eMBB and URLLC slices are 30 and 90, respectively. Each rectangular block in the bar chart represents the RBs allocated to a user. For example, there are a total of eight users in the eMBB slice for the traditional method when the number of users is 30.}
\label{NetSlice}
\end{figure*}
We consider a network consisting of $120$ users with user ID ranging from 1 to 120, uniformly distributed in a $(450\times 450)$ m$^2$ area. A WirelessAgent is deployed at the base station located in the center of this area. The users arrive randomly, and only one user requests service from the base station at each round. According to the user requirement, the WirelessAgent must allocate resources to support the user's service, i.e., network slicing management. The objective of WirelessAgent is to minimize the resource occupation rate by autonomously performing network slicing management based on user requirements and network conditions. We consider two types of slices (i.e., URLLC and eMBB), where the URLLC and eMBB slices have $30$ and $90$ resource blocks (RBs), respectively. Each RB can support different sub-carries and power levels. We assume that each RB can support a data rate of 1 Mb/s. The decision ranges for the URLLC and eMBB slices are $[1, 5]$ Mb/s and  $[5, 20]$ Mb/s, respectively. For example, if a service requires $10$ Mb/s, it will be assigned to the eMBB slice at the user registration step. There is an overlap between the decision ranges, allowing users to be assigned to either the eMBB or URLLC slices based on the slice management algorithm.

This experiment is conducted within a dialog environment. We utilize the application programming interface (API) of the GPT-4o-128K language model as a core component for managing network slices in the WirelessAgent. An example of the WirelessAgent's implementations is illustrated in Fig. \ref{Fig:ImpleDetails}, and more details can be seen at https://github.com/weiiguo/Wireless-Agent. We design the main prompts to guide the WirelessAgent's workflow, which decomposes this task into multiple sub-tasks using the CoT technique. It can be seen that WirelessAgent leverages in-context learning, reflection, and tool manipulation capabilities to complete different sub-tasks. Upon the arrival of a new user, the WirelessAgent initiates the resource allocation task. The planning process aligns closely with the steps outlined in the planning module in Fig. \ref{Case_Study}. This experiment is conducted sequentially until the final user is accommodated.

We compare the WirelessAgent-enabled slice management approach with the traditional slice management method described in \cite{zhang2017network}. In the traditional method, the user's slice type is known prior, and RBs are randomly assigned within the corresponding slice's decision range. Fig. \ref{NetSlice} shows the resource occupation rates of the traditional and the WirelessAgent-enabled slice management approaches, respectively. There are three key observations. First, the WirelessAgent maintains a lower resource occupancy rate than the traditional method. For example, the traditional approach consumes resources number of RBs by $60\%$ to support $10$ users, while the WirelessAgent only consumes resources by $53.3\%$. Second, the WirelessAgent can support more users than the traditional method by dynamically adjusting each user's RBs. For instance, the WirelessAgent supports up to $15$ users in the eMBB slice when the number of users is $30$, while the traditional method supports only $8$ users. Third, the WirelessAgent improves network performance by performing the slice handover algorithm. For example, when the total number of users is 25, and this slice's RBs are exhausted, WirelessAgent reallocates users by moving users (i.e., Users 110 and 115 in Fig. \ref{Case_Study}) in the overlapping area from the eMBB slice to the URLLC slice. 
These results demonstrate that the WirelessAgent can intelligently and autonomously handle network slicing management tasks.

\section{Conclusions and Future Works}
This article introduced WirelessAgent, a novel framework leveraging the power of LLMs to develop AI agents capable of autonomously managing complex tasks in wireless networks. It consists of four core modules: perception, memory, planning, and action, which were substantiated through a case study of network slicing management. Simulation results demonstrated that WirelessAgent surpasses traditional methods in accurately understanding user intent, effectively allocating slice resources, and consistently maintaining optimal performance even under high user loads. 

Several promising directions exist for WirelessAgent:

\textbf{Enhanced multimodal integration:} Further research is needed to develop more sophisticated methods for integrating multimodal data into LLMs. This includes exploring advanced encoding techniques and fusion strategies to fully capture the rich information in wireless environments.

\textbf{Explainable wireless AI:} As agents for wireless networks become involved in critical network management decisions, ensuring transparency and explainability in their decision making process is crucial. Future work should focus on developing methods for interpreting and explaining the reasoning behind agent actions.

\textbf{Security and privacy considerations:} The deployment of agents in real-world wireless networks necessitates robust security and privacy safeguards. Research efforts should address potential vulnerabilities and develop mechanisms to protect sensitive network information and user data.

\textbf{Real-world deployment and evaluation:} Moving beyond simulations, future work should focus on deploying and evaluating agents in real-world wireless networks. This will involve addressing practical challenges related to scalability, robustness, and integration with existing network infrastructure.
\bibliographystyle{ieeetr}
\bibliography{ref}

\noindent{\bf{Jingwen Tong}} [S'19-M'23] (eejwentong@ust.hk) received his Ph.D. at Xiamen University. He currently works as a Postdoc Fellow at HKUST. \\
{\bf{Jiawei Shao}} [S'20-M'24] (jiawei.shao@connect.ust.hk) received his Ph.D. at HKUST, where he currently works as a Postdoc Fellow. \\
{\bf{Qiong Wu}} [S'18-M'23] (cseqiongwu@ust.hk) received her Ph.D. from Sun Yat-sen University. She currently works as a Postdoc Fellow at HKUST.\\
{\bf{Wei Guo}} [S'18-M'24] (eeweiguo@ust.hk) received his Ph.D. from the CUHK-SZ. He currently works as a Postdoc Fellow at HKUST.\\
{\bf{Zijian Li}} (zijian.li@connect.ust.hk) is
currently pursuing his Ph.D. at HKUST. \\
{\bf{Zehong Lin}} [S’17-M’22] (eezhlin@ust.hk) received his Ph.D. degree from CUHK. He is currently a Research Assistant Professor at HKUST. \\
{\bf{Jun Zhang}} [S'06-M'10-SM'15-F'21] (eejzhang@ust.hk) received his Ph.D. from the University of Texas at Austin. He is currently an Associate Professor at HKUST and a Fellow of IEEE.

\end{document}